\documentclass[intlimits,twoside,a4paper]{article}

\usepackage{amsmath,amssymb}
\usepackage{graphicx}
\usepackage{wrapfig}

\usepackage[T2A]{fontenc}
\usepackage[cp1251]{inputenc}

\usepackage{cmpj2}

    \newcommand{\beq}{\begin{equation}}
    \newcommand{\eeq}{\end{equation}}

    \newcommand\beqa{\begin{eqnarray}}
    \newcommand\eeqa{\end{eqnarray}}
    \newcommand{\nn}{\nonumber\\}
    \newcommand{\la}{\lambda}
    \newcommand{\e}{\eta}
    
    \newcommand{\one}{{(1)}}
    \newcommand{\two}{{(2)}}
    \newcommand{\three}{{(3)}}
    \newcommand{\four}{{(4)}}

\usepackage[usenames]{color}

\issue{2012}{15}{2}{23602}
\doinumber{10.5488/CMP.15.23602}
\title[Piece-wise constant potentials]%
{Rational-function approximation for fluids interacting via piece-wise constant potentials}
\author[A. Santos, S.B. Yuste, M. L\'opez de Haro]{A. Santos\refaddr{label1}, S.B.~Yuste\refaddr{label1},  M. L\'opez de Haro\refaddr{label2}}
\addresses{
\addr{label1} Departamento de F\'{\i}sica, Universidad de
Extremadura, E-06071 Badajoz, Spain
\addr{label2}Centro de Investigaci\'on en Energ\'ia, Universidad Nacional Aut\'onoma de M\'exico (U.N.A.M.), \\ Temixco, Morelos 62580, M{e}xico
}

\date{Received January 13, 2012, in final form March 16, 2012}
\authorcopyright{A. Santos, S.B. Yuste, M. L\'opez de Haro, 2012}

\begin{document}

\maketitle

\begin{abstract}
The structural properties of fluids whose molecules interact via potentials with a hard-core plus
$n$ piece-wise constant sections of different widths and heights are derived using a  (semi-analytical) rational-function approximation method.
The results are illustrated for the cases of a square-shoulder plus square-well potential and a shifted square-well  potential
and compared both with simulation data and with those that follow from the  (numerical) solutions of the Percus-Yevick integral equation.
 \\cavity function, rational-function approximation
\pacs 61.20.Gy, 61.20.Ne, 05.20.Jj, 82.70.Dd

\end{abstract}

\section{Introduction}
\label{sec1}
Simple models of intermolecular interaction have  {proven} to be useful tools in understanding diverse phenomena in real fluids. Among these, the so-called discrete potentials {that} include the popular square-well~\cite{R65,BH67,BH67b,KSL72,HMF76,SHT77,LK78,VMRJM92,LKLLW99,SBC05,LSYS05,LRO06,RPPS08,AF08,GPL09,MRR09,HTS09,ERM09,GSC11} and square-shoulder~\cite{HS70,SH72,LKLLW99,YSH11,GSC11} potentials, or combinations of them~\cite{CMA84,CS84,CSD89,BG99,VBG01,FMSBS01,SBFMS04,MFSBS05, BPGS06,GCC07,CBR07,BFNB08,RPPS08,BNB09,RPSP10,FRT10,HTS11,BOO11}, have received much attention. This is not surprising in view of their relative simplicity and versatility, which allows one to treat a variety of problems including, amongst others, chemical reactions~\cite{CS84}, liquid-liquid transitions~\cite{FMSBS01,SBFMS04,MFSBS05,CBR07}, colloidal interactions~\cite{GCC07}, the anomalous density behaviour of water and supercooled fluids~\cite{BFNB08,BNB09} and the thermodynamic and transport properties of Lennard-Jones fluids~\cite{CMA84,CSD89}.

While the phase diagram and the thermodynamic and transport properties of the discrete-potential fluids have been thoroughly examined, studies of their structural properties, either theoretical or from simulations, are more limited. In fact, for the case where the potential is built as a combination of square wells and shoulders, we are only aware of the rather recent work reported in references~\cite{GCC07,HTS11,BOO11}.

In previous papers~\cite{YS94,YSH11}, following a methodology that, although approximate, has proven successful for many other systems~\cite{HYS08}, the structural properties of the square-well and the square-shoulder fluids were derived. The main aim of this paper is to use a similar methodology, referred to as the  method of rational-function approximation (RFA), to generalize the previous results and derive the structural properties of fluids whose molecules interact via potentials with a hard-core plus piece-wise constant sections of different widths and heights.

{It should be emphasized that the RFA is not a new integral equation but rather an alternative approach to derive the structural properties of fluids in an analytic or semi-analytic way. Instead of proposing a closure relation for the Ornstein-Zernike equation, it deals directly with the radial distribution function in Laplace space and involves some coefficients which are determined by imposing (basic) specific physical conditions (see references~\cite{YSH11,YS94,HYS08} for details).
An interesting feature of the method is that, when applied to hard-sphere systems in odd dimensions~\cite{RS07,RRHS08,RS11,RS11b}, its simplest implementation coincides with the exact solution of the Percus-Yevick (PY) integral equation~\cite{W63,W64,T63,L64,FI81,L84,RHS04,RHS07}. In the case of even dimensions, the mathematical problem becomes much more complicated,  which is connected with the absence of a simple relationship between the Laplace transform (which is a key element of the RFA approach) and the Fourier transform (related to the static structure factor) in even dimensions. In fact, the  PY equation for hard-sphere fluids has no analytical solution in  {those} dimensions.}

The paper is organized as follows. In section~\ref{sec2} we provide the background material required for the subsequent development. This is followed in section~\ref{sec3} by two different proposals for the RFA, both for the general $n$-step case and for the particular case of a two-step potential. Section~\ref{results} contains the results of our calculations for illustrative cases and their comparison with both simulation data~\cite{BOO11} and with those that stem out of the numerical solution of the PY equation for this system. We close the paper in section~\ref{concl} with further discussion and some concluding remarks. The consistency of the present approach  {with the results} for both the square-well and square-shoulder results is proven in an appendix.

\section{Fundamental relations}
\label{sec2}
The radial distribution function $g(r)$ of a fluid of particles interacting via a potential $\varphi(r)$ is directly related to the probability of
finding two particles separated by a distance $r$~\cite{HM06}. It can be
measured from neutron or x-ray diffraction
experiments through the static structure factor $S(q)$, which is related to the Fourier transform of $g(r)-1$ by
\begin{eqnarray}
\label{b1}
S(q)&=&1+\rho \int \rd \mathbf{r}\, \re^{-\ri \mathbf{q}\cdot \mathbf{r}} [g(r)-1]\nonumber\\
&=&1-2\pi\rho \left.\frac{G(s)-G(-s)}{s}\right|_{s=\ri q}\, ,
\end{eqnarray}
where  $\rho$ is the
number density and
\begin{equation}
\label{b3}
G(s)=\int_0^\infty \rd r\, \re^{-rs} rg(r)
\end{equation}
is the Laplace transform of $rg(r)$.
The isothermal compressibility of the fluid, $\kappa_T=\rho^{-1}\left(\partial
\rho/\partial p\right)_{T}$, where $p$ is the pressure and $T$ is the temperature, is directly related to the
long-wavelength limit of the structure factor:
\beq
\rho k_{\mathrm{B}}T\kappa_T=S(0),
\label{b2}
\eeq
where $k_{\mathrm{B}}$ is the Boltzmann constant.
Thus, all the physically
relevant  information about the equilibrium state of the fluid is contained in
$g(r)$ or, equivalently, in $G(s)$.

Let us consider now the piece-wise constant potential
\begin{equation}
\label{a1}
\varphi(r)=\left\{
\begin{array}{ll}
\infty  ,& r<\sigma, \\
\epsilon_1  ,& \sigma<r<\lambda_1 \sigma, \\
\epsilon_2  ,& \lambda_1\sigma<r<\lambda_2 \sigma, \\
\vdots&\vdots \\
\epsilon_n  ,& \lambda_{n-1}\sigma<r<\lambda_n \sigma, \\
0,&r>\lambda_n \sigma .
\end{array}
\right.
\end{equation}
This potential is characterized by a hard core of diameter $\sigma$ and $n$ steps of  {(positive or negative)} heights $\epsilon_j$ and widths $(\la_j-\la_{j-1})\sigma$ (where the convention $\la_0=1$ is understood). Thus, $\la_n\sigma$ denotes the total range of $\varphi(r)$. The sign of $\epsilon_j$ defines whether the $j$-th step is either a ``shoulder'' ($\epsilon_j>0$) or a ``well'' ($\epsilon_j<0$). The interaction potential at $r=\la_j\sigma$ ($j=1,\ldots, n$) is repulsive if $\epsilon_j>\epsilon_{j+1}$ and attractive if $\epsilon_j<\epsilon_{j+1}$ (where, by convention, $\epsilon_{n+1}=0$).
As usual, the density is measured by the packing fraction $\eta \equiv \frac{\pi}{6}\rho\sigma^3$.
Henceforth, we will  take the hard-core diameter $\sigma=1$ as the length unit.

It is convenient to define an
auxiliary function $F(s)$ directly related to the Laplace transform $G(s)$ through the relation
\begin{eqnarray}
\label{b5}
G(s)&=&s\frac{F(s)\re^{-s}}{1+12\eta F(s)\re^{-s}}\nonumber\\
 &=&\sum_{m=1}^\infty (-12\eta)^{m-1}s[F(s)]^m \re^{-ms}.
\end{eqnarray}
Laplace inversion of equation (\ref{b5}) provides a useful representation of
the radial distribution function:
 \begin{equation}
 \label{b6}
 g(r)=r^{-1}\sum_{m=1}^\infty (-12\eta)^{m-1}f_m(r-m)\Theta(r-m),
 \end{equation}
where $f_m(r)$ is the inverse Laplace transform of $s[F(s)]^m$ and $\Theta(r)$
is Heaviside's step function. Thus, the knowledge of $g(r)$ [or $G(s)$ or $S(q)$] is fully equivalent to the knowledge of the auxiliary function $F(s)$. The representation~\eqref{b6} reflects the fact that, due to the hard core at $r=1$, the radial distribution function possesses singularities (of decreasing order) at $r=1,2,3,\ldots$.
In particular, the value of $g(r)$ at contact, $g(1^+)$, is given by
the asymptotic behaviour of $F(s)$ for large $s$:
\begin{equation}
\label{b7}
g(1^+)=f_1(0)=\lim_{s\rightarrow \infty} s^2F(s).
\end{equation}
Since $g(1^+)$ should be finite and different from zero, we get the condition
\begin{equation}
\label{b9}
F(s)\sim s^{-2},\quad s\rightarrow \infty.
\end{equation}

{}From equation~\eqref{b1}, it turns out that the behaviour of $G(s)$ for
small $s$
determines the value of $S(0)$:
\begin{equation}
\label{b8}
G(s)=s^{-2}+\text{const}+\frac{1-S(0)}{24\eta}s+\mathcal{O}  (s^2).
\end{equation}
Insertion of equation (\ref{b8}) into the first equality of equation (\ref{b5}) yields
the first five terms in the expansion
of $F(s)$ in
powers of $s$~\cite{YS91,YS94,YHS96},
\begin{equation}
\label{b10}
F(s)=-\frac{1}{12\eta}\left(1+s+\frac{1}{2}s^2+\frac{1+2\eta}{12\eta}s^3
+\frac{2+\eta}{24\eta}s^4\right)+{\cal O}(s^5),
\end{equation}
and expresses $S(0)$ in terms of the coefficients of $s^5$ and $s^6$, namely
\beq
S(0)=\frac{24}{5}\eta^3\left(6\left.\frac{\rd^5F(s)}{\rd s^5}\right|_{s=0} -
\left.\frac{\rd^6F(s)}{\rd s^6}\right|_{s=0}\right)-1+8\eta+2\eta^2.
\label{b4}
\eeq
Equations~\eqref{b9} and~\eqref{b10} provide the behaviours for large $s$ and small $s$, respectively, that the auxiliary function $F(s)$ should necessarily satisfy.

Let us now decompose $F(s)$ as
\beq
F(s)=\sum_{j=0}^n R_j(s)\re^{-(\lambda_j-1)s},
\label{c0}
\eeq
where, as said before, $\lambda_0=1$. The aim of this decomposition is to reflect the discontinuities of $g(r)$ at the points $r=\la_j$ where the potential is discontinuous. Let us denote by $\xi_j(r)$ the inverse Laplace transform
of $s R_j(s)$. Thus,
\beq
\xi_j(0)=\lim_{s\to\infty}s^2R_j(s),\qquad \xi_j'(0)=\lim_{s\to\infty}s\left[
s^2R_j(s)-\xi_j(0)\right],\qquad j=0,\ldots,n.
\label{c3}
\eeq
Equation~\eqref{c0} implies that the inverse Laplace transform of $sF(s)$ is
\beq
f_1(r)=\sum_{j=0}^n \xi_j(r-\lambda_j+1)\Theta(r-\lambda_j+1).
\label{c1}
\eeq

{}From now on we will assume, for the sake of concreteness, that $\lambda_n\leqslant 2$ (although the case $\lambda_n> 2$ can also be dealt with in a similar way). Therefore, equation~\eqref{b6} yields
\beqa
\Theta(2-r)g(r)&=&r^{-1}f_1(r-1)\Theta(r-1)\Theta(2-r)\nn
&=&r^{-1}\sum_{j=0}^n
\xi_j(r-\lambda_j)\Theta(r-\lambda_j)\Theta(2-r).
\label{c2}
\eeqa
As a consequence,
\beq
g(\la_j^+)-g(\la_j^-)=\frac{1}{\la_j}\xi_j(0),\qquad g'(\la_j^+)-g'(\la_j^-)=\frac{1}{\la_j}\left[\xi_j'(0)-\frac{1}{\la_j}\xi_j(0)\right],\qquad j=0,\ldots,n.
\label{disc2}
\eeq
On the other hand,  the cavity function
$y(r)\equiv g(r)\re^{\varphi(r)/k_{\mathrm{B}}T}$ and its first derivative $y'(r)$ are continuous at $r=\la_j$~\cite{A00}. This means that
\beq
g(\la_j^-)\re^{\beta\epsilon_j}=g(\la_j^+)\re^{\beta\epsilon_{j+1}},\qquad g'(\la_j^-)\re^{\beta\epsilon_j}=g'(\la_j^+)\re^{\beta\epsilon_{j+1}},\qquad
j=1,\ldots,n,
\label{c14}
\eeq
or, according to equation~\eqref{c2},
\beq
\xi_j(0)=\left[\re^{\beta\left(\epsilon_{j}-\epsilon_{j+1}\right)}-1\right]
\sum_{i=0}^{j-1}\xi_i(\la_j-\la_i) ,\quad
j=1,\ldots,n,
\label{c15_b}
\eeq
\beq
\xi_j'(0)=\left[\re^{\beta\left(\epsilon_{j}-\epsilon_{j+1}\right)}-1\right]
\sum_{i=0}^{j-1}\xi_i'(\la_j-\la_i) ,\quad
j=1,\ldots,n.
\label{c15_f}
\eeq

In the low-density limit, one has $g(r)\to \re^{-\beta\varphi(r)}$, so that
\beqa
\lim_{\eta\to 0}F(s)&=&s^{-1}\re^s \lim_{\eta\to 0}G(s)\nn
&=&s^{-3}\sum_{j=0}^{n}
\re^{-(\lambda_{j}-1)s}\left(1+\lambda_{j}s\right)\left(\re^{-\beta\epsilon_{j+1}}
-\re^{-\beta\epsilon_j}\right),
\label{c4}
\eeqa
where, apart from $\epsilon_{n+1}=0$ and $\lambda_0=1$, it is understood that $\epsilon_0=\infty$.
Comparison with equation~\eqref{c0} yields
\beq
\lim_{\eta\to
0}R_j(s)=s^{-3}\left(1+\lambda_{j}s\right)\left(\re^{-\beta\epsilon_{j+1}}-\re^{-\beta\epsilon_{j}}\right),\quad j=0,\ldots,n.
\label{c5}
\eeq

\section{Construction of the rational-function approximation}
\label{sec3}
\subsection{The general $n$-step case}
According to the scheme presented in the preceding section, the full knowledge of the radial distribution function $g(r)$ for the $n$-step potential~\eqref{a1} amounts to a prescription  {for} the functions $R_j(s)$, such that the function $F(s)$ obtained through equation~\eqref{c0} fulfills the conditions~\eqref{b9} and~\eqref{b10}.

Now we assume the following \emph{rational-function} approximation (RFA) for $R_j(s)$:
\beq
R_j(s)=-\frac{1}{12\eta}\frac{A_j+B_j s}{1+S_1 s+S_2 s^2+S_3 s^3}, \qquad j=0,\ldots,n.
\label{c6}
\eeq
Since the degree difference between the numerator and denominator of $R_0(s)$ is equal to 2, the form~\eqref{c6} ensures the consistency with equation~\eqref{b9}. In fact, according to equation~\eqref{c3},
\beq
\xi_j(0)=-\frac{1}{12\eta}\frac{B_j}{S_3},\qquad \xi_j'(0)=-\frac{1}{12\eta}\left(\frac{A_j}{S_3}-\frac{B_j S_2}{S_3^2}\right),\qquad j=0,\ldots,n.
\label{Xi0}
\eeq
Insertion into equation~\eqref{disc2} yields
\beq
g(\la_j^+)-g(\la_j^-)=\frac{1}{12\eta\la_j}\frac{B_j}{S_3},\qquad j=0,\ldots,n,
\label{Xi1}
\eeq
\beq
g'(\la_j^+)-g'(\la_j^-)=-\frac{1}{12\eta\la_j}\left(\frac{A_j}{S_3}-\frac{B_j S_2}{S_3^2}-\frac{B_j}{S_3\la_j}\right),\qquad j=0,\ldots,n.
\label{Xi1b}
\eeq
For $r>0$, application of the residue theorem provides the inverse Laplace transform of $sR_j(s)$ as
\beq
\xi_j(r)=-\frac{1}{12\eta}\sum_{\alpha=1}^3 \frac{A_j+B_j s_\alpha}{S_1 +2S_2 s_\alpha+3S_3 s_\alpha^2}s_\alpha \re^{s_\alpha r},
\label{xi}
\eeq
where $s_\alpha$ ($\alpha=1,2,3$) are the three roots of the cubic equation $1+S_1 s_\alpha+S_2 s_\alpha^2+S_3 s_\alpha^3=0$.

The approximation~\eqref{c6} contains $2(n+1)+3=2n+5$ parameters to be determined.
The exact expansion (\ref{b10}) imposes five constraints among them:
\beq
A_0=1-\sum_{j=1}^n A_j,
\label{c7}
\eeq
\beq
S_1=B_0-\left(1+C^\one\right),
\label{c8}
\eeq
\beq
S_2=\frac{1}{2}\left(1+2C^\one+C^\two\right)-B_0,
\label{c9}
\eeq
\beq
S_3=\frac{1}{2}B_0-\frac{1}{12\e}-\frac{1}{6}\left(1+3C^\one+3C^\two+C^\three\right),
\label{c10}
\eeq
\beq
(1+2\e)B_0=1+C^\one+\frac{\e}{2}\left(1+4C^\one+6C^\two+4C^\three+C^\four\right),
\label{c11}
\eeq
where
\beq
C^{(k)}\equiv \sum_{j=1}^n \left[A_j (\lambda_j-1)^k-k
B_j(\lambda_j-1)^{k-1}\right].
\label{c12}
\eeq
Thus, the five coefficients $A_0$, $B_0$, $S_1$, $S_2$ and $S_3$ are given by equations~\eqref{c7}--\eqref{c12} as linear combinations of $A_j$ and $B_j$ ($j=1,\ldots, n$). In the case of a hard-sphere system (formally, $n=0$) the problem is closed and one recovers~\cite{YS91,YHS96,HYS08} the exact solution of the PY integral equation~\cite{T63,W63,W64}. On the other hand, $2n$ parameters remain to be determined if $n\geqslant 1$.

Before dealing with this problem, let us see that the proposal~\eqref{c6} is consistent with the exact form~\eqref{c5} in the zero-density limit.
This implies that one should have
\beq
\lim_{\eta\to 0}S_1= 0,\qquad
\lim_{\eta\to 0} \eta S_2=0,  \qquad \lim_{\eta\to 0}\eta S_3= -\frac{1}{12},
\label{S1}
\eeq
\beq
\lim_{\eta\to 0}A_0=\re^{-\beta\epsilon_{1}},
\label{c13_0}
\eeq
\beq
\lim_{\eta\to 0}B_0=\re^{-\beta\epsilon_{1}},
\label{c13b_0}
\eeq
\beq
\lim_{\eta\to
0}A_j=\re^{-\beta\epsilon_{j+1}}-\re^{-\beta\epsilon_{j}},\qquad j=1,\ldots,n,
\label{c13}
\eeq
\beq
 \lim_{\eta\to 0}{B_j}=\la_j \left(\re^{-\beta\epsilon_{j+1}}-\re^{-\beta\epsilon_{j}}\right),\qquad j=1,\ldots,n.
\label{c13b}
\eeq
First, note that, if equation~\eqref{c13} is satisfied, then equation~\eqref{c7} reduces to equation~\eqref{c13_0}. Next, equations~\eqref{c13} and~\eqref{c13b} imply that $\lim_{\eta\to 0}C^\one=A_0-1$. Thus, equation~\eqref{c11} reduces to equation~\eqref{c13b_0}. Finally, taking the limit $\eta\to 0$ in equations~\eqref{c8}--\eqref{c10} one obtains equation~\eqref{S1}.
This proves that the approximate form~\eqref{c6} becomes exact in the limit $\eta\to 0$, provided that the $2n$ coefficients $A_j$ and $B_j$ ($j=1,\ldots, n$) are consistent with equations~\eqref{c13} and~\eqref{c13b}.

\subsubsection{RFA1}
As a first proposal  {to determine the $2n$ coefficients $A_j$ and $B_j$ with $j=1,\ldots,n$}, we may discard the density dependence of $A_j$  ($j=1,\ldots, n$). As a consequence, those coefficients are fixed at their zero-density values, namely
\beq
A_j=\re^{-\beta\epsilon_{j+1}}-\re^{-\beta\epsilon_{j}},\qquad j=1,\ldots,n.
\label{cc13}
\eeq
Insertion of equation~\eqref{cc13} into equation~\eqref{c7} implies that $A_0$ is also given by its zero-density expression, equation~\eqref{c13_0}.
The coefficients $B_j$  ($j=1,\ldots, n$) are determined from the $n$ constraints~\eqref{c15_b} stemming from the continuity of $y(r)$. Making use of equations~\eqref{Xi0} and~\eqref{xi} yields
\beq
\frac{B_j}{S_3}=\left[\re^{\beta(\epsilon_{j}-\epsilon_{j+1})}-1\right]
\sum_{\alpha=1}^3\frac{s_\alpha \re^{\la_j s_\alpha}}{S_1 +2S_2 s_\alpha+3S_3 s_\alpha^2}\sum_{i=0}^{j-1}(A_i+B_is_\alpha)\re^{-\la_i s_\alpha},\qquad
j=1,\ldots,n.
\label{c15_c}
\eeq
These conditions should be enforced numerically.  {We will refer to this variant as the RFA1 method.} In the one-step case ($n=1$),  {it} coincides with the one previously proposed for square-well~\cite{YS94,AS01,LSAS03,LSYS05,MYS06,HYS08} and square-shoulder~\cite{YSH11} fluids. The RFA1 reduces to the
PY solution for hard spheres and sticky hard spheres in the appropriate limits~\cite{YS94,YSH11}.

\subsubsection{RFA2}
A more sophisticated version  {(here referred to as RFA2)} consists in replacing the simple prescription~\eqref{cc13} by the  enforcement of the continuity of $y'(r)$ via equation~\eqref{c15_f}. Again, taking into account equations~\eqref{Xi0} and~\eqref{xi}, one gets
\beq
\frac{A_j}{S_3}-\frac{B_j S_2}{S_3^2}=\left[\re^{\beta(\epsilon_{j}-\epsilon_{j+1})}-1\right]
\sum_{\alpha=1}^3\frac{s_\alpha^2 \re^{\la_j s_\alpha}}{S_1 +2S_2 s_\alpha+3S_3 s_\alpha^2}\sum_{i=0}^{j-1}(A_i+B_is_\alpha)\re^{-\la_i s_\alpha},\qquad
j=1,\ldots,n.
\label{c15_g}
\eeq
Now the problem requires the numerical solution of  {the} $2n$ coupled transcendental equations~\eqref{c15_c} and \eqref{c15_g}, instead of the $n$ equations~\eqref{c15_c} required in the RFA1.

While the RFA2 is internally more consistent than the RFA1, it is not necessarily more accurate. As we will illustrate in section~\ref{results}, the requirement of a rather subtle continuity condition of $y'(r)$  at $r=\la_j$ may force a radial distribution function with features less realistic than those found.

In either case, once the Laplace transform $G(s)$ is fully determined, its numerical Laplace inversion~\cite{AW92} yields $g(r)$.

\subsection{Particularization to two-step potentials}
Let us now adapt the previous scheme to the case $n=2$.  {Then},
\beq
F(s)=R_0(s)+R_1(s)\re^{-(\lambda_1-1)s}+R_2(s)\re^{-(\lambda_2-1)s},
\label{3.1}
\eeq
where the three functions $R_0(s)$, $R_1(s)$ and $R_2(s)$ have the rational-function form~\eqref{c6}. There are 9 coefficients to be determined. In the simpler RFA1 approach, $A_j$ are fixed by their zero-density values, i.e., according to equation~\eqref{cc13},
\beq
A_0=\re^{-\beta\epsilon_1},\qquad A_1=\re^{-\beta\epsilon_2}-\re^{-\beta\epsilon_1},\qquad A_2=1-e^{-\beta\epsilon_2}.
\label{3.2}
\eeq
Next, $S_1$, $S_2$ and $S_3$ are given as linear combinations of  $B_j$ by equations~\eqref{c8}--\eqref{c10}, where
\beq
C^{(k)}=A_1(\lambda_1-1)^k+A_2(\lambda_2-1)^k-kB_1(\lambda_1-1)^{k-1}-kB_2(\lambda_2-1)^{k-1}.
\label{3.4}
\eeq
Moreover, equation~\eqref{c11}  expresses  $B_0$ in terms $B_1$ and $B_2$. Thus, only two equations are needed. They are given by equation~\eqref{c15_c}, i.e.,
\beq
\frac{B_1}{S_3}=\left[\re^{\beta(\epsilon_{1}-\epsilon_2)}-1\right]
\sum_{\alpha=1}^3\frac{s_\alpha \re^{(\la_1-1) s_\alpha}}{S_1 +2S_2 s_\alpha+3S_3 s_\alpha^2}(A_0+B_0s_\alpha),
\label{3.5}
\eeq
\beq
\frac{B_2}{S_3}=\left(\re^{\beta\epsilon_{2}}-1\right)\sum_{\alpha=1}^3\frac{s_\alpha \re^{\la_2 s_\alpha}}{S_1 +2S_2 s_\alpha+3S_3 s_\alpha^2}\left[(A_0+B_0s_\alpha)\re^{-s_\alpha}+(A_1+B_1s_\alpha)\re^{-\la_1 s_\alpha}\right].
\label{3.6}
\eeq
These two transcendental equations close the problem.

In a more sophisticated RFA2 version of the approximation, equation~\eqref{3.2} is replaced by
\beq
A_0=1-A_1-A_2,
\label{3.7}
\eeq
and equation~\eqref{c15_g}, i.e.,
\beq
\frac{A_1}{S_3}-\frac{B_1 S_2}{S_3^2}=\left[\re^{\beta(\epsilon_{1}-\epsilon_2)}-1\right]
\sum_{\alpha=1}^3\frac{s_\alpha^2 \re^{(\la_1-1) s_\alpha}}{S_1 +2S_2 s_\alpha+3S_3 s_\alpha^2}(A_0+B_0s_\alpha),
\label{3.8}
\eeq
\beq
\frac{A_2}{S_3}-\frac{B_2 S_2}{S_3^2}=\left(\re^{\beta\epsilon_{2}}-1\right)\sum_{\alpha=1}^3\frac{s_\alpha^2 \re^{\la_2 s_\alpha}}{S_1 +2S_2 s_\alpha+3S_3 s_\alpha^2}\left[(A_0+B_0s_\alpha)\re^{-s_\alpha}+(A_1+B_1s_\alpha)\re^{-\la_1 s_\alpha}\right].
\label{3.9}
\eeq
These two equations should be solved in conjunction with equations~\eqref{3.5} and~\eqref{3.6}.

 {It is shown in the appendix}  that the RFA for the two-step potential reduces to the one for the one-step potential in the limit $\epsilon_1=\epsilon_2$ as well as in the limit $\epsilon_1\to\infty$.

\section{Illustration of the approximation for two-step potentials}
\label{results}

In this section we illustrate our proposals RFA1 and RFA2 in two representative examples of two-step potentials. First, we consider a square-shoulder plus square-well (SS+SW) potential (i.e., $\epsilon_1>0$ and $\epsilon_2<0$) with $\epsilon_1=|\epsilon_2|$, $\lambda_1=1.25$ and $\lambda_2=1.5$. The second case corresponds to a shifted square-well (sSW) potential (i.e., $\epsilon_1=0$ and $\epsilon_2<0$) with  $\lambda_1=1.25$ and $\lambda_2=1.5$. Both potentials have recently been analyzed by B\'arcenas et al.~\cite{BOO11} by means of exchange replica Monte Carlo (MC) simulations, with special emphasis on the  coexistence curves between the vapour and the condensed phase (liquid or solid) and the structural properties of the condensed phase at coexistence.

The chosen state points are $\rho\sigma^3=0.421$, $k_{\mathrm{B}}T/|\epsilon_2|=0.64$ for the SS+SW system and $\rho\sigma^3=0.427$, $k_{\mathrm{B}}T/|\epsilon_2|=0.76$ for the sSW system, both lying on the condensed branch of the respective coexistence curve~\cite{BOO11}. The radial distribution function $g(r)$ at a subcritical temperature and at the density of the coexisting liquid represents a rather stringent test of the theoretical approach developed in this paper. To put the results in a proper context, we have also numerically solved the PY integral equation by an iterative method~\cite{YSH11}. The comparison between RFA and PY is especially relevant since both theories coincide in the hard-sphere and sticky-hard-sphere limits~\cite{YS94,YSH11}.


\begin{figure}[ht]
\centerline{\includegraphics[width=0.36\textwidth]{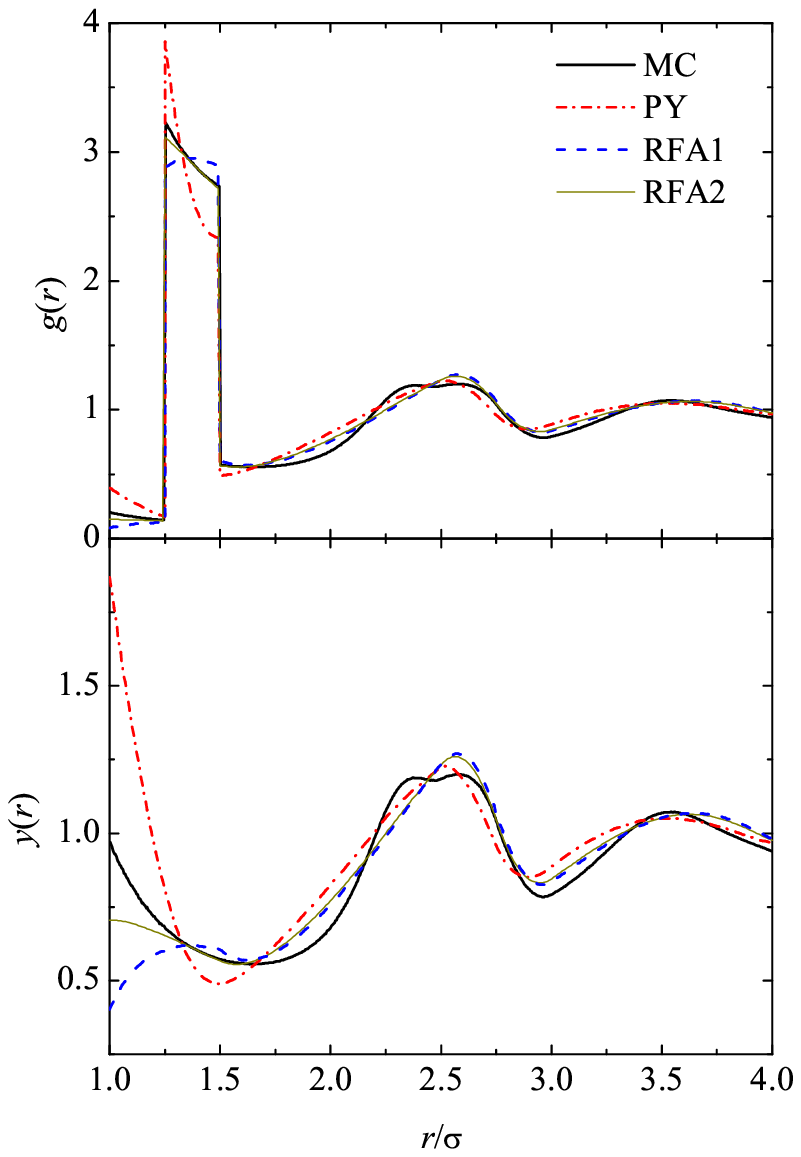}
\hspace{2cm}
\includegraphics[width=0.36\textwidth]{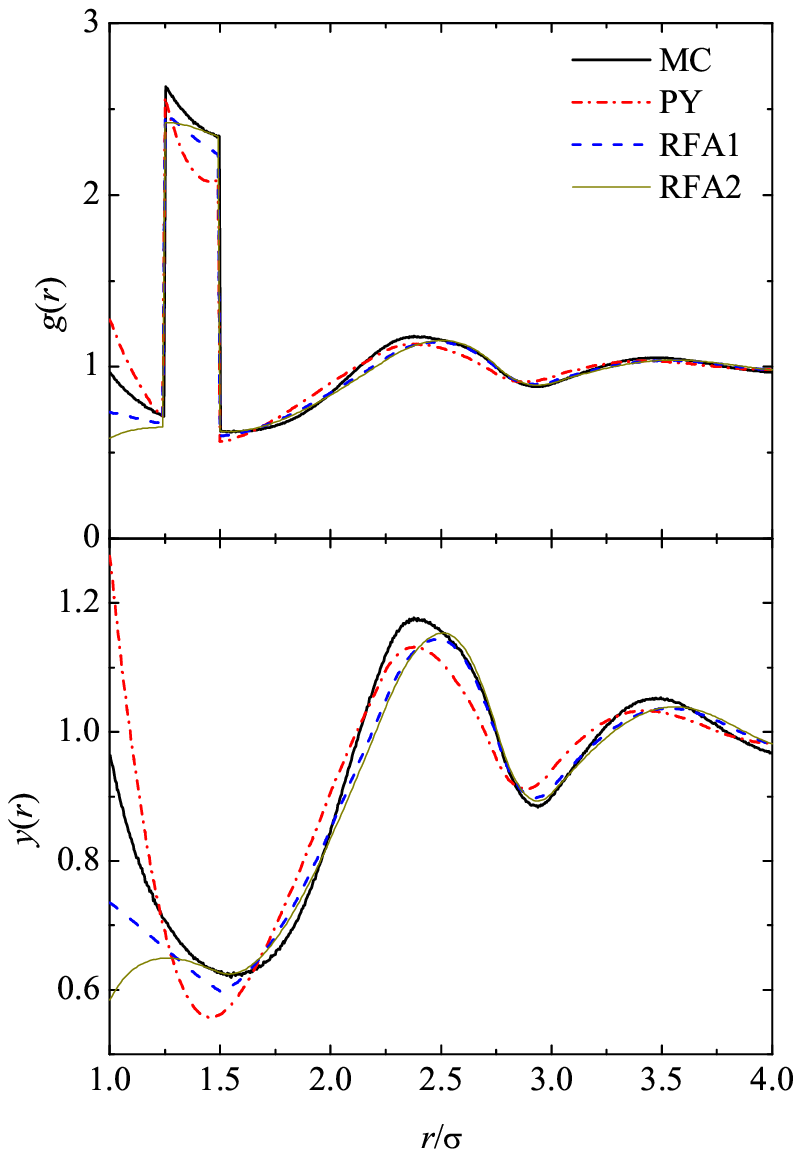}}
\centerline{
\parbox[t]{0.48\textwidth}{\caption{(Color online) Radial distribution function $g(r)$ (top panel) and cavity function $y(r)$ (bottom panel) for a square-shoulder plus square-well potential ($\epsilon_1>0$ and $\epsilon_2<0$) with $\epsilon_1=|\epsilon_2|$, $\lambda_1=1.25$ and $\lambda_2=1.5$ at $\rho\sigma^3=0.421$ and $k_{\mathrm{B}}T/|\epsilon_2|=0.64$. The thick solid lines are MC results~\cite{BOO11}, while the dashed-dotted, dashed and thin solid lines correspond to the PY, RFA1 and RFA2 theories, respectively.} \label{fig1}}
\hfill
\parbox[t]{0.48\textwidth}{\caption{(Color online) Radial distribution function $g(r)$ (top panel) and cavity function $y(r)$ (bottom panel) for a shifted square-well potential ($\epsilon_1=0$ and $\epsilon_2<0$) with $\lambda_1=1.25$ and $\lambda_2=1.5$ at $\rho\sigma^3=0.427$ and $k_{\mathrm{B}}T/|\epsilon_2|=0.76$. The thick solid lines are MC results~\cite{BOO11}, while the dashed-dotted, dashed and thin solid lines correspond to the PY, RFA1 and RFA2 theories, respectively.} \label{fig2}}
}
\end{figure}

Figures \ref{fig1} and \ref{fig2} show the radial distribution function and the cavity function for the SS+SW and sSW systems, respectively, as obtained from MC simulations~\cite{BOO11}, the PY numerical solution and our RFA1 and RFA2 approaches.
We observe from the figures that the RFA1 and RFA2 curves are practically indistinguishable in the region $r>\lambda_2$, where they describe the main oscillating trends of the true distribution, at least in a semi-quantitative way.
Inside the well ($\lambda_1\leqslant r\leqslant \lambda_2$), however, both versions of the RFA differ. In the SS+SW case (figure~\ref{fig1}), the RFA2 prediction is excellent, visibly differing from the MC data only near the inner radius of the well, $r=\lambda_1$, while the RFA1 exhibits an accused convex shape; as a consequence, in the shoulder region ($1\leqslant r\leqslant\la_1$) the RFA1 predicts a $g(r)$ with positive slope. In the sSW case (figure~\ref{fig2}), on the other hand, it is the RFA2 that presents a convex shape inside the well ($\lambda_1\leqslant r\leqslant \lambda_2$) and hence exhibits a positive slope in the shift region ($1\leqslant r\leqslant\la_1$). By contrast, the RFA1, which (as clearly seen in the bottom panel of figure~\ref{fig2}) has a discontinuous first derivative at $r=\la_2$, presents an almost linear  {form} in the regions  $1\leqslant r\leqslant\la_1$ and $\lambda_1\leqslant r\leqslant \lambda_2$ with a qualitatively correct behaviour. Note, however, that the RFA2 is very accurate in the region inside the well and near the outer radius ($1.4\leqslant r\leqslant \la_2$).

As for the PY theory, it is obvious from figures~\ref{fig1} and \ref{fig2} that it performs worse than any of the RFA results for $r\geqslant \la_1$ and than either RFA2 (SS+SW system) or RFA1 (sSW system) for $1\leqslant r\leqslant \la_1$. This  {remarkable} result (given that the PY solution requires full numerical work, in contrast to the semi-analytical character of the RFA) was already observed for square-shoulder fluids~\cite{YSH11}.

\section{Conclusions}
\label{concl}
In summary, in this paper we have  {proposed a} method of deriving the structural properties of a particular kind of discrete-potential fluids, namely the ones in which molecules interact via a potential consisting of a hard-core plus $n$-step constant sections of different heights and widths.  {The method is based on assuming  rational-function forms for $n+1$ functions $R_j(s)$ defined in Laplace space, the coefficients being constrained by physical consistency conditions.} Two approximations were considered: one in which some unknown constants are fixed at their zero density value (RFA1) and one which enforces the continuity of the derivative of the cavity function (RFA2). In the former case, one has in the end to solve (numerically) $n$ coupled transcendental equations while in the latter, which is internally more consistent, the set to be solved is made of $2n$ coupled transcendental equations.

The illustrative cases of the SS+SW   {and  sSW potentials} that we examined indicate that our approach, being semi-analytical in nature,  {is  a} reasonable compromise between simplicity and accuracy. For the SS+SW potential, the RFA2 approximation gives the best results, whereas the RFA1 is superior in the sSW potential. On the other hand, in these illustrative cases the present development also represents a clear improvement over the solutions of the corresponding  PY integral equations, which require harder numerical labor.

These further examples of the adequacy of the RFA method as compared with the results of the PY equation reinforce the notion that such methodology is a valuable alternative to the integral equation approach for the derivation of the structural properties of fluid systems.
 {This conclusion is of course based on a limited analysis. One could easily argue that the comparison with the results stemming out of the Ornstein-Zernike equation with either the Rogers--Young or the hypernetted chain closures would be a more solid test ground. However, these approximations require much harder numerical work,  {while our approach yields the explicit $s$-dependence of the Laplace transform $G(s)$}. In any case, we find it interesting to compare the performance of the present development with the results of some other simple  approximations such as the hybrid mean spherical approximation considered in reference~\cite{GCC07} or the first-order mean spherical approximation used in reference~\cite{HTS11}. We plan to undertake such comparisons in our future work.}

\section*{Acknowledgements}
This paper, dealing with a subject in which he has made significant contributions, is dedicated to Orest Pizio on the occasion of his sixtieth birthday. We want to thank G.~Odriozola for kindly supplying all the simulation data of reference~\cite{BOO11}. Two of us (A.S. and S.B.Y) acknowledge the financial support of the  {Spanish Government} through Grant No.~FIS2010--16587 and  the Junta de Extremadura (Spain) through Grant No.~GR10158 (partially financed by FEDER funds). The work of M.L.H. has been partially supported by DGAPA-UNAM under project IN--107010--2.

\appendix

\section{Consistency with the one-step (square-well or square-shoulder) case}
\label{app}

Let us denote by $g(r;\epsilon_1/\epsilon_2,\lambda_1,\lambda_2,T^*,\eta)$ the radial distribution function in the case of a two-step potential with reduced temperature $T^*=k_{\mathrm{B}}T/\epsilon_2$ and packing fraction $\eta=\frac{\pi}{6}\rho$, where the hard-core distance is $\sigma=1$. Analogously, we denote by $\bar{g}(r;\lambda,T^*,\eta)$ the radial distribution function of a one-step  {(square-well or square-shoulder)} potential of unit hard-core distance and relative width $\lambda$ with reduced temperature $T^*$ (in units of the step height) and packing fraction $\eta$.

The two-step problem should reduce to the  {one-step problem}  in certain limits. The aim of this appendix is to prove that the RFA2 method fulfills these consistency conditions. A similar and simpler proof can be worked out for the RFA1 method.

\subsection{Case $\epsilon_1=\epsilon_2$}

An obvious physical condition is that the two-step system becomes equivalent to a one-step system in the limit $\epsilon_1\to\epsilon_2$, i.e.,
\beq
g(r;1,\lambda_1,\lambda_2,T^*,\eta)=\bar{g}(r;\lambda_2,T^*,\eta).
\label{A1}
\eeq
 In Laplace space,
\beq
G(s;1,\lambda_1,\lambda_2,T^*,\eta)=\bar{G}(s;\lambda_2,T^*,\eta),
\label{A2}
\eeq
\beq
F(s;1,\lambda_1,\lambda_2,T^*,\eta)=\bar{F}(s;\lambda_2,T^*,\eta).
\label{A3}
\eeq
Taking into account  {equations~\eqref{c0} and~\eqref{c6}}, one should have
\begin{align}
A_0(1,\lambda_1,\lambda_2,T^*,\eta)&=\bar{A}_0(\lambda_2,T^*,{\eta}),
\label{A4}\\
A_1(1,\lambda_1,\lambda_2,T^*,\eta)&=0,
\label{A5}\\
A_2(1,\lambda_1,\lambda_2,T^*,\eta)&=\bar{A}_1(\lambda_2,T^*,{\eta}),
\label{A6}\\
B_0(1,\lambda_1,\lambda_2,T^*,\eta)&=\bar{B}_0(\lambda_2,T^*,{\eta}),
\label{A7}\\
B_1(1,\lambda_1,\lambda_2,T^*,\eta)&=0,
\label{A8}\\
B_2(1,\lambda_1,\lambda_2,T^*,\eta)&=\bar{B}_1(\lambda_2,T^*,{\eta}),
\label{A9}\\
S_1(1,\lambda_1,\lambda_2,T^*,\eta)&=\bar{S}_1(\lambda_2,T^*,{\eta}),
\label{A10}\\
S_2(1,\lambda_1,\lambda_2,T^*,\eta)&=\bar{S}_2(\lambda_2,T^*,{\eta}),
\label{A11}\\
S_3(\infty,\lambda_1,\lambda_2,T^*,\eta)&=\bar{S}_3(\lambda_2,T^*,{\eta}).
\label{A12}
\end{align}
Equations~\eqref{A4}--\eqref{A6} are in full agreement with equation  {\eqref{c7}}. It remains to prove that the above equations  are consistent with equations~\eqref{c8}--\eqref{c11},~\eqref{3.5},~\eqref{3.6},~\eqref{3.8} and~\eqref{3.9}.

{}For simplicity, henceforth we will omit the arguments of the barred and unbarred quantities. {}From equation~\eqref{3.4},
\beq
C^{(k)}=\bar{C}^{(k)}=\bar{A}_1(\lambda_2-1)^k-k\bar{B}_1(\lambda_2-1)^{k-1},
\label{A14}
\eeq
where use has been made of equations~\eqref{A5}--\eqref{A9}. This ensures that equations~\eqref{c8}--\eqref{c11} are satisfied by the one-step system, provided they are satisfied by the two-step system in the limit $\epsilon_1\to\epsilon_2$. Next, the transcendental equations~\eqref{3.5} and~\eqref{3.8} are trivially satisfied by equations~\eqref{A5} and~\eqref{A8}. Finally, the transcendental equations~\eqref{3.6} and~\eqref{3.9} become the same for the two- and one-step cases.

\subsection{Case $\epsilon_1\to\infty$}
In the limit $\epsilon_1\to\infty$ the two-step potential becomes equivalent to a one-step potential of hard-core distance $\lambda_1$ and relative width $\lambda\equiv\lambda_2/\lambda_1$ with a packing fraction $\bar{\eta}\equiv\eta\lambda_1^3$, i.e.,
\beq
g(r;\infty,\lambda_1,\lambda_2,T^*,\eta)=\bar{g}(r/\lambda_1;\lambda,T^*,\bar{\eta}).
\label{B1}
\eeq
In Laplace space, equation~\eqref{B1} translates into
\beq
G(s;\infty,\lambda_1,\lambda_2,T^*,\eta)=\lambda_1^2\bar{G}(\lambda_1 s;\lambda,T^*,\bar{\eta}).
\label{B2}
\eeq
{}From equation~\eqref{b5},
\beq
F(s;\infty,\lambda_1,\lambda_2,T^*,\eta)=\lambda_1^3\re^{-(\lambda_1-1)s}\bar{F}(\lambda_1 s;\lambda,T^*,\bar{\eta}).
\label{B3}
\eeq
According to  {equations~\eqref{c0} and~\eqref{c6}}, one should have
\begin{align}
A_0(\infty,\lambda_1,\lambda_2,T^*,\eta)&=0,
\label{B4}\\
A_1(\infty,\lambda_1,\lambda_2,T^*,\eta)&=\bar{A}_0(\lambda,T^*,\bar{\eta}),
\label{B5}\\
A_2(\infty,\lambda_1,\lambda_2,T^*,\eta)&=\bar{A}_1(\lambda,T^*,\bar{\eta}),
\label{B6}\\
B_0(\infty,\lambda_1,\lambda_2,T^*,\eta)&=0,
\label{B7}\\
B_1(\infty,\lambda_1,\lambda_2,T^*,\eta)&=\lambda_1\bar{B}_0(\lambda,T^*,\bar{\eta}),
\label{B8}\\
B_2(\infty,\lambda_1,\lambda_2,T^*,\eta)&=\lambda_1\bar{B}_1(\lambda,T^*,\bar{\eta}),
\label{B9}\\
S_1(\infty,\lambda_1,\lambda_2,T^*,\eta)&=\lambda_1\bar{S}_1(\lambda,T^*,\bar{\eta}),
\label{B10}\\
S_2(\infty,\lambda_1,\lambda_2,T^*,\eta)&=\lambda_1^2\bar{S}_2(\lambda,T^*,\bar{\eta}),
\label{B11}\\
S_3(\infty,\lambda_1,\lambda_2,T^*,\eta)&=\lambda_1^3\bar{S}_3(\lambda,T^*,\bar{\eta}).
\label{B12}
\end{align}
Again, equations~\eqref{B4}--\eqref{B6} are consistent with equation   {\eqref{c7}}. Let us now check that equations~\eqref{B4}--\eqref{B12} are consistent with equations~\eqref{c8}--\eqref{c11},~\eqref{3.5},~\eqref{3.6},~\eqref{3.8} and~\eqref{3.9}.

{}As before, for simplicity we omit the arguments of the barred and unbarred quantities. Taking into account the definition~\eqref{3.4}, one has
\beq
\bar{C}^{(k)}=\bar{A}_1(\lambda-1)^k-k\bar{B}_1(\lambda-1)^{k-1}.
\label{B14}
\eeq
According to equations~\eqref{B4}--\eqref{B9}, we get that, in the limit $\epsilon_1\to\infty$,
\beq
C^{(k)}=(\lambda_1-1)^k+\bar{A}_1\left[(\lambda_2-1)^k-(\lambda_1-1)^k\right]-k\bar{B}_0\lambda_1(\lambda_1-1)^{k-1}-k\bar{B}_1\lambda_1(\lambda_2-1)^{k-1},
\label{B15}
\eeq
where we have made use of the property $\bar{A}_0+\bar{A}_1=1$.
{}From equations~\eqref{B14} and~\eqref{B15} one has
\beq
1+C^\one=\lambda_1\left(1+\bar{C}^\one-\bar{B}_0\right),\ \
\label{B16}
\eeq
\beq
1+2C^\one+C^\two=\lambda_1^2\left(1+2\bar{C}^\one+\bar{C}^\two-2\bar{B}_0\right),
\label{B17}
\eeq
\beq
1+3C^\one+3C^\two+C^\three=\lambda_1^3\left(1+3\bar{C}^\one+3\bar{C}^\two+\bar{C}^\three-3\bar{B}_0\right),
\label{B18}
\eeq
\beq
1+4C^\one+6C^\two+4C^\three+C^\four=\lambda_1^4\left(1+4\bar{C}^\one+6\bar{C}^\two+4\bar{C}^\three+\bar{C}^\four-4\bar{B}_0\right).
\label{B19}
\eeq
Equations~\eqref{B16}--\eqref{B18}, together with equations~\eqref{c8}--\eqref{c10}, prove the consistency of equations~\eqref{B10}--\eqref{B12}, respectively.
Next, equations~\eqref{B16} and~\eqref{B19} prove that, if~\eqref{c11} is satisfied by the unbarred parameters, so it is by the barred parameters.

Now we consider the transcendental equations that need to be additionally solved. In the case of the one-step potential, equations~\eqref{c15_c} and~\eqref{c15_g} reduce to
\beq
\frac{\bar{B}_1}{\bar{S}_3}=\left(\re^{1/T^*}-1\right)\sum_{\alpha=1}^3\frac{\bar{s}_\alpha \re^{(\la-1) \bar{s}_\alpha}}{\bar{S}_1 +2\bar{S}_2 \bar{s}_\alpha+3\bar{S}_3 \bar{s}_\alpha^2}(\bar{A}_0+\bar{B}_0\bar{s}_\alpha),
\label{B20}
\eeq
\beq
\frac{\bar{A}_1}{\bar{S}_3}-\frac{\bar{B}_1 \bar{S}_2}{\bar{S}_3^2}=\left(\re^{1/T^*}-1\right)\sum_{\alpha=1}^3\frac{\bar{s}_\alpha^2 \re^{(\la-1) \bar{s}_\alpha}}{\bar{S}_1 +2\bar{S}_2 \bar{s}_\alpha+3\bar{S}_3 \bar{s}_\alpha^2}(\bar{A}_0+\bar{B}_0\bar{s}_\alpha),
\label{B20.2}
\eeq
respectively, where $\bar{s}_\alpha$ ($\alpha=1,2,3$) are the three roots of the cubic equation $1+\bar{S}_1 \bar{s}_\alpha+\bar{S}_2 \bar{s}_\alpha^2+\bar{S}_3 \bar{s}_\alpha^3=0$.
In the case of the two-step potential, multiplication of both sides of equations~\eqref{3.5} and~\eqref{3.8} by $\re^{-\beta\epsilon_1}$ shows that they are identically satisfied in the limit $\epsilon_1\to\infty$ if $A_0=B_0=0$ [cf. equations~\eqref{B4} and~\eqref{B7}]. As for equations~\eqref{3.6} and~\eqref{3.9}, they become
\beq
\frac{B_2}{S_3}=\left(\re^{1/T^*}-1\right)\sum_{\alpha=1}^3\frac{s_\alpha \re^{(\la_2-\la_1) s_\alpha}}{S_1 +2S_2 s_\alpha+3S_3 s_\alpha^2}(A_1+B_1s_\alpha),
\label{B21}
\eeq
\beq
\frac{A_2}{S_3}-\frac{B_2S_2}{S_3^2}=\left(\re^{1/T^*}-1\right)\sum_{\alpha=1}^3\frac{s_\alpha^2 \re^{(\la_2-\la_1) s_\alpha}}{S_1 +2S_2 s_\alpha+3S_3 s_\alpha^2}(A_1+B_1s_\alpha).
\label{B21.2}
\eeq
Taking into account the equations~\eqref{B5},~\eqref{B6} and~\eqref{B8}--\eqref{B12}, as well as the fact that $\bar{s}_\alpha=\la_1 s_\alpha$, it is straightforward to check that equations~\eqref{B20} and~\eqref{B20.2} are equivalent to equations~\eqref{B21} and~\eqref{B21.2}, respectively. This closes the proof of equation~\eqref{B1}.



\vspace{-7mm}

\ukrainianpart

\title%
{Наближення раціональними функціями для плинів, що взаємодіють через відтинково-сталі потенціали}
\author[]{А. Сантос\refaddr{label1},  С.Б. Юсте\refaddr{label1},
 М. Лопес де Гаро\refaddr{label2}}
\addresses{
\addr{label1} Фізичний факультет, Університет Екстремадури, E-06071
Бадайос, Іспанія
\addr{label2}Центр енергетичних досліджень,
Національний автономний університет Мексики (U.N.A.M.), \\ Морельос
62580, Мексика }

\makeukrtitle

\begin{abstract}
\tolerance=3000%
Використовуючи (напіваналітичний) метод наближення
раціональними функціями, отримано структурні властивості плинів, молекули яких взаємодіють через
потенціали із твердою серцевиною та $n$ сходинками різної ширини та
висоти. Результати ілюструються на прикладах
потенціалу прямокутний уступ плюс прямокутна яма і потенціалу
зміщена прямокутна яма, і порівнюються з результатами симуляцій, а
також з результатами, що отримуються з (числових) розв'язків
інтегрального рівняння Перкуса-Євіка.

\keywords дискретні потенціали, модель прямокутного уступу, модель прямокутної ями, радіальна функція розподілу, функція порожнини,
наближення раціональними функціями
\end{abstract}

\end{document}